\documentclass[sigconf]{acmart}
\AtBeginDocument{%
  }

\copyrightyear{2025}
\acmYear{2025}
\setcopyright{cc}
\setcctype{by}
\acmConference[SIGIR-AP 2025]{Proceedings of the 2025 Annual International ACM SIGIR Conference on Research and Development in Information Retrieval in the Asia Pacific Region}{December 7--10, 2025}{Xi'an, China}
\acmBooktitle{Proceedings of the 2025 Annual International ACM SIGIR Conference on Research and Development in Information Retrieval in the Asia Pacific Region (SIGIR-AP 2025), December 7--10, 2025, Xi'an, China}\acmDOI{10.1145/3767695.3769505}
\acmISBN{979-8-4007-2218-9/2025/12}

\usepackage{multirow}
\usepackage{acronym}
\usepackage{tcolorbox}
\acrodef{IR}{information retrieval}
\acrodef{LLM}{large language model}
\acrodef{NLP}{natural language processing} 
\acrodef{RAG}{Retrieval-Augmented Generation}
\acrodef{NLI}{Natural Language Inference}

\begin{document}
\title{VeriCite: Towards Reliable Citations in Retrieval-Augmented Generation via Rigorous Verification}

\author{Haosheng Qian, Yixing Fan}
\authornote{Corresponding author.}
\affiliation{%
  \institution{State Key Laboratory of AI Safety, ICT, CAS \\ University of Chinese Academy of Sciences, CAS}
  \city{Beijing}
  \country{China}
}
\email{qianhaosheng22@mails.ucas.ac.cn}
\email{fanyixing@ict.ac.cn}

\author{Jiafeng Guo}
\affiliation{%
  \institution{State Key Laboratory of AI Safety, ICT, CAS \\ University of Chinese Academy of Sciences, CAS}
  \city{Beijing}
  \country{China}
}
\email{guojiafeng@ict.ac.cn}

\author{Ruqing Zhang}
\affiliation{%
  \institution{State Key Laboratory of AI Safety, ICT, CAS \\ University of Chinese Academy of Sciences, CAS}
  \city{Beijing}
  \country{China}
}
\email{zhangruqing@ict.ac.cn}

\author{Qi Chen}
\affiliation{%
  \institution{Meituan Inc.}
  \city{Beijing}
  \country{China}
}
\email{cqict90@gmail.com}

\author{Dawei Yin}
\affiliation{%
  \institution{Baidu Inc.}
  \city{Beijing}
  \country{China}
}
\email{yindawei@acm.org}

\author{Xueqi Cheng}
\affiliation{%
  \institution{State Key Laboratory of AI Safety, ICT, CAS \\ University of Chinese Academy of Sciences, CAS}
  \city{Beijing}
  \country{China}
}
\email{cxq@ict.ac.cn}

\renewcommand{\shortauthors}{Haosheng et al.}

\begin{abstract}
\Ac{RAG} has emerged as a crucial approach for enhancing the responses of \acp{LLM} with external knowledge sources. Despite the impressive performance in complex question-answering tasks, \ac{RAG} still struggles with hallucinations. Attributing RAG-generated content through in-line citations has demonstrated potential in reducing hallucinations and facilitating human verification. Existing citation generation methods primarily rely on either fine-tuning the generator or employing post-processing approaches for citation matching. However, the former approach demands substantial annotated data and computational resources, while the latter often encounters difficulties in managing multiple citations and frequently produces suboptimal results. In this paper, we introduce a novel framework, called \textbf{VeriCite}, designed to rigorously validate supporting evidence and enhance answer attribution. Specifically, \textbf{VeriCite} breaks down into a three-stage generation: 1) The \textit{initial answer generation} first generates a response based on all available contexts and has its claims verified through the NLI model; 2) the \textit{supporting evidence selection} assesses the utility of each document and extracts useful supporting evidences; 3) the \textit{final answer refinement} integrates the initial response and collected evidences to produce the final, refined answer. We conduct experiments across five open-source \acp{LLM} and four datasets, demonstrating that VeriCite can significantly improve citation quality while maintaining the correctness of the answers. \footnote{Our code is publicly available at \url{https://github.com/QianHaosheng/VeriCite}}
\end{abstract}



\maketitle

\section{Introduction}
Retrieval-Augmented Generation (RAG) \cite{lewis2020retrieval, shuster2021retrieval, gao2023retrieval} plays a crucial role in enabling large language models (LLMs) \cite{brown2020language, liu2024deepseek} to tackle challenges such as real-time news queries and domain-specific issues, thereby expanding the capabilities and application scope of \acp{LLM}.
However, as retrieval technology is not always flawless, it simultaneously introduces new challenges for \acp{LLM}.
For example, if irrelevant information is retrieved and used as a reference, the \ac{LLM} may incorporate this noise and generate incorrect answers, exacerbating the hallucination issue\cite{filippova2020controlled, ji2023survey, thakur2023nomiracl}.

Therefore, enabling \acp{LLM} to generate attributable responses is vital for ensuring trustworthiness and mitigating misinformation. 
An effective strategy to enhance the reliability of LLM responses is through citation mechanisms, whereby each statement is explicitly anchored to relevant source materials \cite{gao2023enabling, li2024improving, fan2025trustrag}.
This approach not only establishes traceability by allowing users to independently verify the accuracy of responses, but also facilitates error diagnosis and promotes transparency in human-AI collaboration.

Current approaches for generating answers with citations can be broadly classified into two paradigms  \cite{huang2023citation}.
The first category, classified as ``intrinsic attribution'', operates synchronously with text generation. 
These approaches typically treat citations as regular tokens and enable LLM to directly generate citations within answers through fine-tuning or in-context learning \cite{nakano2021webgpt, liu2023webglm, gao2023enabling}.
Nevertheless, intrinsic integration approaches face several practical constraints:
(1) Fine-tuning demands extensive domain-specific annotation and significant computational resources;
(2) In-context learning is highly sensitive to the input examples, leading to poor generalization performance.

The other category can be classified as ``extrinsic attribution'', which initially generates a draft answer and subsequently employs post-processing approaches to match retrieved passages with statements in the answer.
Common matching methods include utilizing sentence similarity metrics such as BLEU \cite{papineni2002bleu} and ROUGE \cite{lin2004rouge}, or employing \ac{NLI} classifiers to evaluate entailment relationships \cite{gao2023rarr}. 
Classic similarity metrics are computationally efficient, but their effectiveness is constrained by the challenge of determining thresholds.
Conversely, although \ac{NLI} models deliver higher accuracy, yet fundamentally struggle to handle cases where a single statement requires multiple citations \cite{gao2023enabling}.

\begin{figure*}[t]
  \centering
  \includegraphics[width=1\linewidth]{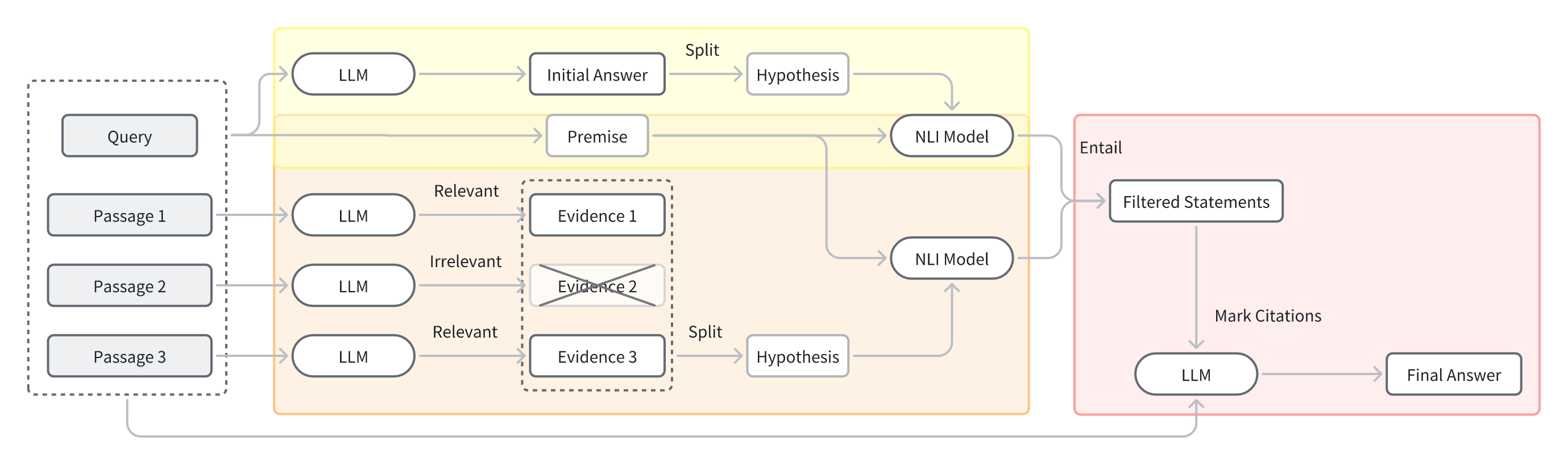}
  \caption{Overview of VeriCite framework.}
  \label{figure1}
\end{figure*}

To address the aforementioned issues, we propose a novel framework named \textbf{VeriCite}, which strengthens the reliability of citations through rigorous verification. 
In contrast to previous studies which primarily focused on the answer generation process or post-processing stages, VeriCite concentrates on the phase after retrieved passages are obtained but before final answer generation commences.
VeriCite consists of three stages: initial answer generation, supporting evidence selection, and final answer refinement (as illustrated in Figure~\ref{figure1}).
The initial answer generation stage generates a response based on all retrieval passages and uses an \ac{NLI} model to verify the citations in the statement, ensuring the reliability of the answer.
The subsequent supporting evidence selection stage thoroughly extracts potentially useful evidence from each passage. 
This evidence must also undergo verification through the \ac{NLI} model, and the verified evidence is then marked with citations.
The final answer refinement stage integrates the initial answer and the extracted evidence, with the \ac{LLM} responsible for reorganizing the order of the statements to improve fluency, removing redundant content, and merging citations.

VeriCite aims to pre-screen the content within retrieved passages that is genuinely valuable for answer generation, pre-attributing citations to these high-quality segments to ensure source traceability. 
This preprocessing approach helps eliminate noise from the input, significantly alleviating the cognitive load on \acp{LLM} when extracting key information from long contexts. 
Furthermore, the strategy of pre-attributing citations reduces the model's attribution difficulty, enabling the generator to more seamlessly and accurately reuse existing citations within the answer. 
Extensive experiments conducted across multiple datasets and LLMs demonstrate that while achieving answer accuracy on par with baselines, VeriCite yields a significant improvement in citation generation quality.

\section{Related Work}
In retrieval-augmented question answering, methods for generating attributed answers typically fall into two primary categories.

The first category employs ``intrinsic attribution'', leveraging generative models' inherent attribution capabilities.
This approach typically utilizes supervised fine-tuning or in-context learning to enable models to produce answers with integrated citations.
Among seminal implementations, WebGPT \cite{nakano2021webgpt} enhances open-domain question answering (QA) accuracy by simulating human web browsing behavior.
Built upon GPT-3 \cite{brown2020language}, this system extracts relevant webpage passages as supporting evidence and inserts citations as commandS within answers.
The authors trained reward models on extensive human preference annotations, optimizing answer quality through Proximal Policy Optimization \cite{schulman2017proximal}.
Subsequent innovations include WebGLM \cite{liu2023webglm}, which integrates an LLM-augmented retriever, bootstrapped generator, and a human preference-aware scorer.
Its automated annotation pipeline enabled large-scale training data generation, with supervised fine-tuning on citation-annotated QA data yielding robust attribution capabilities.
APO \cite{li2024improving} advances training methodology by formulating attribution as preference learning and introducing a progressive optimization framework with sentence-level rewards that enhances alignment efficiency.
Distinctively, LongCite \cite{zhang2024longcite} tackles fine-grained citation in long-context QA through its Coarse to Fine (CoF) data construction scheme, enabling precise sentence-level attribution with superior traceability relative to passage-level alternatives.
Unlike approaches requiring fine-tuning, alternative approaches employ prompting to instruct models to incorporate citations during answer generation.
ALCE \cite{gao2023enabling} systematically evaluated multiple few-shot citation generation strategies, including Vanilla, Summary, and Snippet.
Alternative research efforts have designed more sophisticated reasoning pipelines dedicated to enabling models to perform proactive verification and citation refinement during generation. 
For instance, VTG \cite{sun2024towards} introduces a document storage mechanism with long short-term memory, implements an active retrieval component that generates diversified queries, and incorporates a hierarchical verification module featuring an evidence finder to validate relationships between generated answers and their citations.

Contrastingly, ``extrinsic attribution'' methods incorporate citations during post-processing.
These methods first generate an initial answer (with or without citations) using a generative model, then establish correspondences between the generated text and retrieved passages through text matching techniques, and finally insert appropriate citations~\cite{10.1145/3331184.3331403}.
This strategy enables attribution even for models lacking inherent citation capabilities. 
For instance, WebGLM's automated citation annotation pipeline utilized the ROUGE-1 \cite{lin2004rouge} similarity metric to evaluate citation correctness, filtering higher-quality training data.
Beyond text similarity metrics, alternative approaches leverage \ac{NLI} models to determine entailment relationships between answer sentences and retrieved passages, assigning citations based on classification results. 
ALCE implemented this NLI approach as a representative post-processing baseline method.

Effective evaluation methodologies are indispensable for advancing citation generation research, with established approaches encompassing both human assessment and automated metrics~\cite{press2024citeme,10.1145/2766462.2767710}.
The pioneering human evaluation framework, Attributable to Identified Sources (AIS) \cite{rashkin2023measuring}, measures textual faithfulness to source materials through a structured protocol: 
annotators first examine model-generated text to determine whether each statement requires external source substantiation, then verify (1) the presence of explicit source attribution, and (2) content consistency between generated claims and corresponding source materials.
While human evaluation offers superior accuracy, its significant drawbacks include high labor costs and low efficiency. 
To address these challenges, researchers proposed AutoAIS \cite{gao2023rarr} based on AIS, which leverages \ac{NLI} models to approximate human judgment.
This automated approach refines evaluation granularity to the sentence level by examining entailment relationships between responses and source materials.
Building upon this foundation, ALCE redefines citation recall and citation precision metrics while establishing the first benchmark for \ac{LLM} attribution evaluation. 
This benchmark incorporates multi-dimensional evaluation of fluency, correctness, and citation quality.
Further advancing the field, CAQA \cite{hu2024benchmarking} introduces a comprehensive four-category framework (Supported, Insufficient, Contradictory, Irrelevant) for fine-grained attribution evaluation, enabling more precise quantification of attribution performance.

\section{VeriCite framework}

\subsection{Task Formulation}
Following previous work \cite{gao2023enabling, liu2023webglm}, the formal description of this task is as follows: Given a query \( q \), top-\( k \) passages \( P = \{ p_1, p_2, \dots, p_k \} \) are retrieved, and the \ac{LLM} needs to generate an answer \( A \). The answer \( A \) consists of several statements \( A = \{ s_1, s_2, \dots \} \). Each statement \( s_i \) may cite a set of passages \( C_i = \{ c_{i1}, c_{i2}, \dots \} \), where \( c_{ij} \in [1, k] \cap \mathbb{Z} \).

\subsection{Initial Answer Generation}
The initial answer generation phase follows standard RAG methodology \cite{lewis2020retrieval}, where the query \( q \) and all top-$k$ retrieved passages \( \{ p_1, p_2, \dots, p_k \} \) are concatenated into a single input sequence for the \ac{LLM}, producing an initial answer \( init\_ans \).

\begin{equation}
  init\_ans = Answer(q, p_1, p_2, \dots, p_k)
\end{equation}

At this stage, we employ a few-shot instruction template to provide in-context learning examples, guiding the \ac{LLM} to learn the citation patterns demonstrated in the exemplars. 
This approach explicitly requires the model to incorporate citations within each answer statement.
The instruction template at this stage is shown in Appendix~\ref{A:1}.
This phase aims to produce a foundational answer for subsequent refinement, requiring only citation incorporation without additional model constraints. 
While contemporary \acp{LLM} excel at answering simple commonsense queries, they inevitably exhibit hallucination tendencies when confronted with complex problems. 

To enhance answer reliability, we implement a rigorous verification and filtering mechanism. 
During initial answer validation, unsupported content must be systematically eliminated, retaining exclusively evidence-substantiated answer statements.
To facilitate granular reliability verification, the initial answer \( init\_ans \) is decomposed into a set of statements \( \{s_1, s_2, \dots\} \), where each statement \( s_i \) is associated with a potentially empty set of citations \( \{c_{i1}, c_{i2}, \dots\} \).
The verification process employs a \ac{NLI} model \( \phi \) trained for Recognizing Textual Entailment (RTE) tasks \cite{dagan2015recognizing, yin2021docnli}, which predicts whether a hypothesis is entailed by, contradicts, or is neutral to given premises.
Specifically, for each statement \( s_i \), we validate whether the corresponding retrieval passages \( \{p_{c_{ij}}\} \) (premises) entail the statement \( s_i \) (hypothesis).

\begin{equation}
  sup_{i} = \phi(concat(p_{c_{ij}}), s_{i})
\end{equation}

The verification outcome \( sup_{i} \) is binary-valued: 
when the model determines that the answer statement \( s_i \) is entailed by (a combination of) the retrieved passages, \( sup_{i} \) is assigned \textit{True}; 
otherwise, it returns \textit{False}, indicating an unsupported statement likely containing hallucinated content that consequently fails verification and must be discarded.

\subsection{Supporting Evidence Selection} 

\begin{table*}[h]
  \centering
      \begin{tabular}{clcccc}
        \hline
         - & Settings & ASQA & ELI5 & HotpotQA & MuSiQue \\
        \hline
        \multirow{3}{*}{Dataset statistics} & Task & Long-form QA & Long-form QA & Multihop QA & Multihop QA \\ \cline{2-6}
        & Question Type & Factoid & How/Why/What & Factoid & Factoid \\
        \cline{2-6}
        & \# Examples & 948 & 1000 & 500 & 500 \\
        
        \hline
        \multirow{2}{*}{ \textit{Evaluation metrics} } & Correctness & EM Recall & Claim Recall & EM Recall & EM Recall \\
        \cline{2-6}
        & Citation Quality &  \multicolumn{4}{c}{ Citation Recall, Citation Precision, Citation F1 }\\
        \hline
      \end{tabular}
  \caption{Statistics of different datasets.}
  \label{tab4}
\end{table*}

Irrelevant information can interfere with the \ac{LLM}'s response generation, potentially causing critical relevant details to be overlooked.
While conventional \ac{RAG} approaches generate answers based on coarsely aggregated retrieval results, our methodology additionally incorporates fine-grained evidence extraction.
This necessity arises from the fundamental misalignment between retriever and generator objectives in standard \ac{RAG} pipelines \citep{liu2024invar}.
These distinct models often exhibit mutually incompatible relevance judgments.
Retrievers may introduce either irrelevant information or seemingly relevant but non-actionable content, both of which can cause generators to overlook genuinely critical information while producing noise-degraded outputs.

Inspired by recent studies \cite{sachan2022improving, qin2024large, sun2023chatgpt}, our evidence selection phase leverages the \ac{LLM}'s robust natural language understanding capabilities to collaborate with the retriever in context extraction.
This dual engagement strategy comprehensively excavates potentially valuable content that might otherwise be overlooked within each passage.
Specifically, the \ac{LLM} first independently evaluates each retrieved passage's utility for answering the query using the instruction template depicted in Appendix~\ref{A:2}.

\begin{equation}
rel_i = Check(q, p_i)
\end{equation}

Following the \ac{LLM}'s secondary verification of passage utility, retained passages (\(rel_i=True\)) proceed to evidence selection.
The generator then independently produces answers for each qualifying passage \(p_i\) using the instruction template shown in Appendix~\ref{A:3}.

\begin{equation}
evidence_i = \left\{
    \begin{array}{ll}
    Answer(q, p_i) & ,rel_i=True \\
    None & ,rel_i=False
    \end{array}
\right.
\end{equation}

Like other texts generated by \acp{LLM}, \(evidence_i\) remains prone to hallucination issues, necessitating further verification of its entailment relationship with the corresponding original passage \(p_i\).
Similar to the previous phase, the verification process decomposes \(evidence_i\) into statements  \( \{s_{i1}, s_{i2}, \dots\} \).
We then employ the \ac{NLI} model \(\phi\) to verify whether the original retrieval passage entails these statements.

\begin{equation}
  sup_{ij} = \phi(p_i, s_{ij})
\end{equation}

Statements \(s_{ij}\) verified by the \ac{NLI} model as entailed by passage \(p_i\) are retained for subsequent summarization and automatically annotated with the corresponding citation marker ``[i]''.
This design fundamentally decouples attribution from generation during the summarization phase.
Final answer citations directly reuse these pre-process markers rather than relying on the generator's attribution capabilities, thereby significantly reducing demands on the \ac{LLM}'s citation capacity.


\subsection{Final Answer Refinement}

Following rigorous collection and verification procedures in the preceding stages, we obtain a curated set of semantically validated statements \( \{ s_1, s_2, \dots \} \) accompanied by their corresponding citation sets \( \{ c_1 = \{ c_{11}, c_{12}, \dots \}, c_2 = \{ c_{21}, c_{22}, \dots \}, \dots \} \). 
While these components exhibit high reliability due to rigorous verification, their inherent fragmentation and potential redundancy render them unsuitable for direct concatenation into a coherent final response.
The refinement phase fundamentally redefines the large language model's role rather than directly addressing the query or making attribution decisions, the model now functions as a synthesis engine.
This engine processes the verified statements and citations as foundational input material, executing three critical transformations: restructuring logical flow and sentence sequencing to enhance coherence, eliminating redundant content to improve conciseness, and strategically consolidating citations to optimize referential clarity.

\begin{equation}
  final\_ans = Refine(q, P, s_1, c_1, s_2, c_2, \dots)
\end{equation}

To mitigate potential referential ambiguity and ensure contextual fidelity, the original retrieved passages are incorporated into the model's input stream. 
This architectural choice provides essential grounding context, enabling more accurate interpretation of statement semantics and preventing summarization errors arising from ambiguous references.
Furthermore, explicit instructional constraints mandate that the model preserve the original semantic content of input statements without modification while simultaneously ensuring the final output maintains both informational completeness and fluent logical progression. 
The model must achieve this dual objective through careful rhetorical reorganization rather than content alteration.
The instruction template at this stage is shown in Appendix~\ref{A:4}.

\section{Experiments}
\subsection{Datasets and Models}
To comprehensively evaluate our method's effectiveness across diverse question types, we conduct experiments on four benchmark datasets. 
The long-form QA datasets include ASQA \citep{stelmakh2022asqa}, an ambiguity-aware factual dataset distinguished from conventional benchmarks by its exclusive focus on ambiguous questions sourced from AmbigQA \cite{min2020ambigqa}.
Each query admits multiple valid interpretations, necessitating models to recognize inherent ambiguities and synthesize comprehensive responses using evidentiary support.
Complementing this, ELI5 \citep{fan2019eli5} comprises predominantly non-factual questions originating from Reddit's ``Explain Like I'm Five''\footnote{https://www.reddit.com/r/explainlikeimfive} forum. 
Characterized by complex \textit{how}, \textit{why}, and \textit{what} queries, this dataset presents significant challenges in generating logically coherent, information-rich long-form explanations.
For multi-hop reasoning evaluation, we employ HotpotQA \citep{yang2018hotpotqa} and MuSiQue \citep{trivedi2022musique}. 
HotpotQA features curated factual questions requiring cross-document evidence integration through manually designed multi-step reasoning. 
Conversely, MuSiQue contains synthetically generated factual questions formed by composing single-hop queries, typically demanding 2-4 inference steps. 
This automated composition process yields linguistically structured questions that present heightened analytical difficulty relative to conventional benchmarks.

The ASQA and ELI5 datasets are subsets released by ALCE \citep{gao2023enabling}, while HotpotQA and MusiQue are subsets released by IRCOT \citep{trivedi2023interleaving}. 
Each dataset is evaluated in terms of answer correctness and citation quality. 
Among these, we use the EM (Exact Match) Recall metric to evaluate the answer correctness for the ASQA, HotpotQA, and MusiQue datasets, use Claim Recall to evaluate the answer correctness for the ELI5 dataset, and use Citation F1 to evaluate the citation quality for all datasets.
Dataset details are summarized in Table~\ref{tab4}.

Experiments were conducted on five open-source LLMs: Llama3-8B-Instruct \citep{dubey2024llama}, Gemma-2-9B-it \citep{team2024gemma}, GLM-4-9B-Chat \citep{glm2024chatglm}, Qwen2.5-7B-Instruct \citep{yang2024qwen2}, and Qwen2.5-14B-Instruct.

\begin{table*}
  \centering
  \scalebox{0.95}{
  \begin{tabular}{cccccccccccc}
    \hline
    \multirow{2}{*}{Model} & \multirow{2}{*}{Method} & \multicolumn{2}{c}{ASQA} & \multicolumn{2}{c}{ELI5} & \multicolumn{2}{c}{HotpotQA} & \multicolumn{2}{c}{MuSiQue} & \multicolumn{2}{c}{Overall}\\
    & & EM & Citation F1 & Claim & Citation F1 & EM & Citation F1 & EM & Citation F1 & Correct & Citation F1 \\
    \hline
    \multirow{5}{*}{Llama3-8B} & Vanilla & 38.41 & 69.48 & 12.80 & 38.33 & 43.60 & 35.76 & 12.20 & 17.33 & 26.16 
& 44.35 \\
    & Summary & 37.81 & 65.21 & 10.60 & 42.32 & 36.80 & 24.46 & 3.20 & 4.72 & 22.54 & 40.27 \\
    & Snippet & 35.91 & 57.26 & 11.40 & 38.29 & 41.40 & 24.98 & 10.40 & 15.91 & 24.20 & 38.34 \\
    & APO & 38.12 & 57.73 & \textbf{13.57} & 26.31 & \textbf{46.40} & 37.77 & \textbf{16.20} & 19.38 & \textbf{27.48} & 37.18 \\
    & VeriCite & \textbf{41.63} & \textbf{77.73} & 10.60 & \textbf{59.09} & 42.40 & \textbf{45.72} & 8.40 & \textbf{21.31} & 25.60 & \textbf{56.41} \\
    \hline
    \multirow{5}{*}{Gemma2-9B} & Vanilla & 35.69 & \textbf{77.66} & 11.27 & 43.00 & 40.00 & 45.92 & 6.60 & 15.61 & 23.20 
& 49.99 \\
    & Summary & 36.43 & 72.78 & 8.80 & 40.07 & 41.80 & 44.30 & 8.40 & 15.45 & 23.21 & 47.13 \\
    & Snippet & 34.49 & 69.60 & 10.40 & 41.22 & 38.60 & 43.73 & 7.60 & 19.26 & 22.45 & 47.05 \\
    & APO & 38.18 & 50.58 & \textbf{12.13} & 26.76 & \textbf{49.20} & 32.79 & \textbf{15.40} & 16.59 & \textbf{27.35} & 33.72 \\
    & VeriCite & \textbf{38.93} & 74.89 & 9.90 & \textbf{50.66} & 39.40 & \textbf{63.97} & 7.40 & \textbf{29.70} & 23.81 & \textbf{57.16} \\
    \hline
    \multirow{5}{*}{Glm4-9B} & Vanilla & 38.58 & 69.73 & \textbf{14.40} & 31.24 & \textbf{47.60} & 40.53 & \textbf{14.40} & 23.02 & 27.81 & 43.80 \\
    & Summary & 36.43 & \textbf{72.78} & 11.43 & 34.61 & 34.60 & 22.67 & 6.00 & 14.43 & 22.48 & 41.43 \\
    & Snippet & 35.08 & 66.54 & 12.27 & 31.48 & 30.60 & 18.18 & 5.40 & 8.77 & 21.55 & 36.65 \\
    & APO & 36.83 & 58.74 & 11.33 & 30.35 & 46.00 & 41.26 & 13.80 & 23.93 & 25.83 & 40.24 \\
    & VeriCite & \textbf{43.12} & 71.30 & 12.67 & \textbf{39.66} & 47.00 & \textbf{50.83} & 12.20 & \textbf{27.41} & \textbf{28.20} & \textbf{49.65} \\
    \hline
    \multirow{5}{*}{Qwen2.5-7B} & Vanilla & 37.38 & 70.99 & \textbf{14.00} & 42.71 & 47.60 & 42.62 & 12.60 & 21.75 & 26.98 & 48.24 \\
    & Summary & 37.48 & 70.32 & 12.33 & 39.79 & 38.40 & 24.69 & 7.00 & 9.57 & 23.94 & 41.92 \\
    & Snippet & 35.38 & 68.18 & 12.80 & 36.97 & 32.40 & 19.91 & 4.80 & 6.53 & 22.03 & 38.95 \\
    & APO & 36.53 & 60.69 & \textbf{14.00} & 24.45 & 47.40 & 41.06 & 14.00 & 24.45 & 26.91 & 38.92 \\
    & VeriCite & \textbf{39.47} & \textbf{76.82} & 12.13 & \textbf{55.32} & \textbf{49.40} & \textbf{52.87} & \textbf{14.80} & \textbf{38.87} & \textbf{27.70} & \textbf{59.03} \\
    \hline
    \multirow{5}{*}{Qwen2.5-14B} & Vanilla & 42.03 & 69.49 & \textbf{15.67} & 41.94 & 53.40 & 41.73 & 16.00 & 20.62 & 30.60 & 47.15 \\
    & Summary & 41.64 & 63.38 & 15.10 & 36.74 & 45.20 & 31.24 & 7.00 & 8.58 & 27.37 & 39.60 \\
    & Snippet & 39.29 & 60.83 & 14.37 & 36.91 & 41.00 & 25.66 & 6.40 & 7.43 & 25.55 & 37.70 \\
    & APO & 39.94 & 56.33 & 13.87 & 34.29 & 54.20 & 40.17 & 15.20 & 22.27 & 29.32 & 40.34 \\
    & VeriCite & \textbf{43.50} & \textbf{76.02} & 13.70 & \textbf{56.90} & \textbf{54.40} & \textbf{50.70} & \textbf{16.20} & \textbf{30.77} & \textbf{30.61} & \textbf{57.57} \\
    \hline 
  \end{tabular}}
  \caption{Comparisons between VeriCite and baselines.}
  \label{tab1}
\end{table*}

\subsection{Baselines}
For baseline comparisons, we selected four established approaches: 
\begin{itemize}
    \item \textbf{Vanilla}~\cite{gao2023enabling}: The query and top-\( k \) retrieved passages are concatenated to form the model input. Task-specific instructions coupled with in-context learning mechanisms guide the generation of answers with integrated citations. This approach represents the foundational methodology for attribution generation, processing retrieved passages without additional refinement.
    \item \textbf{Summary}~\cite{gao2023enabling}: Retrieved passages undergo summarization-based compression prior to model input. These summarized compressions are concatenated with the original query and processed through identical task-specific instructions and in-context learning mechanisms to guide the generation of answers with integrated citations. This approach intentionally mitigates textual redundancy in model inputs, enhancing focus on salient information.
    \item \textbf{Snippet}~\cite{gao2023enabling}: Contrasting with the Summary approach, this methodology employs extractive summarization for model input. This methodology preserves exact expressions from retrieved passages, thereby circumventing potential semantic distortion inherent in abstractive summarization.
    \item \textbf{APO}~\cite{li2024improving}: Automatic Preference Optimization framework enhances model performance through a dual-phase approach: supervised fine-tuning followed by preference optimization. During the preference optimization phase, a novel loss function is implemented to enable fine-grained sentence-level rewards, facilitating more efficient model parameter updates.
\end{itemize}

\subsection{Implementation Settings}

In the experiment, the top-\( 5 \) retrieved passages are provided for each query, and each method is given two few-shot examples for in-context learning. 
In the VeriCite method, we use TRUE \citep{honovich2022true} as the NLI model for citation verification.
To ensure the reproducibility of the experiment, all \acp{LLM} generate responses using greedy decoding.

\subsection{Main Results}
Our experimental results, as shown in Table~\ref{tab1}. 

On the ASQA dataset, VeriCite exhibits a clear advantage in answer correctness across all five models, outperforming all baseline methods. 
Notably, the GLM-4 model delivers the most substantial improvement with a 4.54\% increase in correctness over the best performing Vanilla baseline.
Regarding citation quality, Llama3 and Qwen2.5 models achieve significant enhancements in citation F1, surpassing the strongest baseline. In contrast, Gemma-2 and GLM-4 perform marginally below their respective optimal baselines in this metric.

For the ELI5 dataset, VeriCite underperforms relative to the more robust baselines in answer correctness across all five models, indicating potential limitations in its answer generation mechanism for non-factoid questions. 
It is noteworthy that the extensively fine-tuned APO baseline demonstrates strong correctness here, exhibiting only minor degradation with the GLM-4 model. 
Conversely, VeriCite achieves substantial gains in citation quality, with all five models exceeding the best baseline by an average margin of 11.41\% in Citation F1 score, thereby validating its efficacy for citation optimization.

On multi-hop QA datasets, VeriCite shows a pronounced improvement in answer correctness exclusively with the Qwen2.5 model, which surpasses all other baselines. 
However, its performance with the remaining three models falls slightly below their respective best baselines. 
This observation suggests that VeriCite’s supporting evidence selection stage may be suboptimal for multi-hop scenarios requiring cross-passage information integration, highlighting a potential area for architectural refinement. 
Despite this, all models exhibit exceptional citation quality, significantly outperforming the strongest baselines in Citation F1 scores.

Overall, the results indicate that VeriCite matches or exceeds the best baselines in answer correctness, with particularly notable gains observed for the Qwen2.5 and GLM-4 models. Simultaneously, citation quality was significantly enhanced across all five models compared to the best baseline performances. Furthermore, both parameter scales of the Qwen2.5 model exhibited similar improvements, suggesting that the proposed method retains its potential for application to larger-scale models.

\section{Analysis}
\subsection{Ablation Study}

\begin{table}
  \centering
  \begin{tabular}{lcccc}
  \hline
  \multirow{2}{*}{} & Correct & \multicolumn{3}{c}{Citation} \\
  \cline{2-5}
  & EM & Recall & Precision & F1 \\
  \hline
  VeriCite & \textbf{41.63} & \textbf{81.13} & \textbf{74.61} & \textbf{77.73} \\
  -w/o init answer & 39.24 & 76.07 & 71.20 & 73.55 \\
  -w/o evidience selection & 38.57 & 79.42 & 71.82 & 75.43\\
  -w/o NLI verification & 41.59 & 70.99 & 66.95 & 68.91\\
  \hline
  \end{tabular}
  \caption{Ablation study on ASQA.}
  \label{tab2}
\end{table}

This section presents an ablation study conducted on the VeriCite framework to evaluation the contribution of its core components. 
Experiments were performed using the Llama3-8B-Instruct model on the ASQA dataset.

Three specific ablation variants were investigated. 
The first variant omits the initial answer generation stage; consequently, the final answer is generated exclusively utilizing statements derived from the supporting evidence selection stage. 
The second variant removes the supporting evidence selection stage, with the final answer organized solely based on statements obtained from the initial answer generation stage. 
The third variant eliminates the \ac{NLI} based verification module employed in both the initial answer generation and supporting evidence selection stages. 
Under this condition, all generated statements are assumed to be supported by the retrieved passages, effectively bypassing the verification process.

The results detailed in Table~\ref{tab2} reveal significant insights. 
The removal of either the initial answer generation stage or the supporting evidence selection stage induces a substantial decline in answer correctness. 
In contrast, the detrimental effect on citation quality resulting from these omissions is comparatively less pronounced. 
This observation indicates that statements originating from both stages possess a complementary nature, collectively contributing to the comprehensiveness of the final answer. 
Conversely, the ablation of the NLI verification module demonstrates a negligible impact on answer correctness. 
However, this removal causes a severe deterioration in citation quality. 
This finding underscores the critical role of the verification step in ensuring the reliability of the citations within the final answer.

\begin{table}
  \centering
  \begin{tabular}{lcccc}
  \hline
  \multirow{2}{*}{} & Correct & \multicolumn{3}{c}{Citation} \\
  \cline{2-5}
  & EM & Recall & Precision & F1 \\
  \hline
  NLI verifier& 36.88 & \textbf{84.92} & 75.71 & \textbf{80.05} \\
  Llama3-8B verifier & 35.75 & 76.48 & 69.85 & 73.01 \\
  DeepSeek-R1 verifier & \textbf{37.04} & 82.83 & \textbf{75.92} & 79.22 \\
  \hline
  \end{tabular}
  \caption{Results of different verifiers in the VeriCite method.}
  \label{tab3}
\end{table}

\subsection{Discussion of Verifier}
This section discusses the question of verifier model selection within the VeriCite framework.
Recognizing that general \acp{LLM} possess substantial natural language understanding capabilities, employing the same LLM to perform both answer generation and statement verification tasks within VeriCite offers a promising avenue for significantly reducing framework complexity. 
Consequently, we investigate an integrated approach where a single LLM is tasked with generating answers and verifying the support for individual statements within retrieved passages. 
This verification is implemented by instructing the model to output a binary judgment (``Yes’’ or ``No’’) regarding whether each statement is supported by its corresponding passage.
Furthermore, the experimental design incorporates the current SOTA LLM, DeepSeek-R1, specifically for the verification task.
The comparative evaluation utilized the Llama3-8B-Instruct model exclusively for answer generation and evaluated the effectiveness of these three distinct verifier configurations—namely, the LLM verifier, the DeepSeek-R1 verifier, and the NLI verifier—on a randomly selected subset of 200 samples from the ASQA dataset.

Experimental results present in Table~\ref{tab3}. 
Utilizing a general LLM for dual-role verification proved detrimental, leading to a noticeable decline in both answer correctness and citation quality relative to the NLI verifier. 
In contrast, the DeepSeek-R1 verifier achieved a marginal improvement in answer correctness compared to the NLI verifier, while its impact on citation quality was nearly equivalent. 
In terms of computational efficiency, while the LLM and NLI verifiers are comparable in both parameter sizes and operational costs, their practical deployment costs are substantially lower than those of the large-scale DeepSeek-R1 model.

Therefore, based on this empirical evaluation balancing performance gains against resource expenditure, selecting the NLI model for the verification role emerges as the optimal choice, offering an effective and cost-efficient solution.

\section{Conclusion}

In this paper, we propose VeriCite, a novel framework designed to enhance citation quality in \ac{RAG} systems. 
The framework operates through three sequential stages: initial answer generation, supporting evidence selection, and final answer refinement.
Experimental results demonstrate that VeriCite significantly enhances citation quality while maintaining answer correctness comparable to the strongest baseline methods. 
Furthermore, ablation studies confirm the necessity of each core component within the framework. 
Additionally, the paper discusses the critical importance of selecting a \ac{NLI} model for the verification role, providing justification for this design choice.

\begin{acks}
This work was funded by the National Natural Science Foundation of China (NSFC) under Grants No.~62372431, 62472408 and 62441229, the Strategic Priority Research Program of the CAS under Grants No.~XDB0680301, the National Key Research and Development Program of China under Grants No.~2023YFA1011602,
the Lenovo-CAS Joint Lab Youth Scientist Project, and the project under Grants No.~JCKY2022130C039. This work is also supported by Meituan and Baidu Inc.
All content represents the opinion of the authors, which is not necessarily shared or endorsed by their respective employers and/or sponsors.
\end{acks}

\bibliographystyle{ACM-Reference-Format}
\bibliography{custom}


\begin{thebibliography}{42}


\ifx \showCODEN    \undefined \def \showCODEN     #1{\unskip}     \fi
\ifx \showISBNx    \undefined \def \showISBNx     #1{\unskip}     \fi
\ifx \showISBNxiii \undefined \def \showISBNxiii  #1{\unskip}     \fi
\ifx \showISSN     \undefined \def \showISSN      #1{\unskip}     \fi
\ifx \showLCCN     \undefined \def \showLCCN      #1{\unskip}     \fi
\ifx \shownote     \undefined \def \shownote      #1{#1}          \fi
\ifx \showarticletitle \undefined \def \showarticletitle #1{#1}   \fi
\ifx \showURL      \undefined \def \showURL       {\relax}        \fi
\providecommand\bibfield[2]{#2}
\providecommand\bibinfo[2]{#2}
\providecommand\natexlab[1]{#1}
\providecommand\showeprint[2][]{arXiv:#2}

\bibitem[Brown et~al\mbox{.}(2020)]%
        {brown2020language}
\bibfield{author}{\bibinfo{person}{Tom Brown}, \bibinfo{person}{Benjamin Mann},
  \bibinfo{person}{Nick Ryder}, \bibinfo{person}{Melanie Subbiah},
  \bibinfo{person}{Jared~D Kaplan}, \bibinfo{person}{Prafulla Dhariwal},
  \bibinfo{person}{Arvind Neelakantan}, \bibinfo{person}{Pranav Shyam},
  \bibinfo{person}{Girish Sastry}, \bibinfo{person}{Amanda Askell},
  {et~al\mbox{.}}} \bibinfo{year}{2020}\natexlab{}.
\newblock \showarticletitle{Language models are few-shot learners}.
\newblock \bibinfo{journal}{\emph{Advances in neural information processing
  systems}}  \bibinfo{volume}{33} (\bibinfo{year}{2020}),
  \bibinfo{pages}{1877--1901}.
\newblock


\bibitem[Dagan et~al\mbox{.}(2015)]%
        {dagan2015recognizing}
\bibfield{author}{\bibinfo{person}{Ido Dagan}, \bibinfo{person}{Dan Roth},
  \bibinfo{person}{Mark Sammons}, {and} \bibinfo{person}{Fabio~Massimo
  Zanzotto}.} \bibinfo{year}{2015}\natexlab{}.
\newblock \showarticletitle{Recognizing Textual Entailment: Models and
  Applications}.
\newblock \bibinfo{journal}{\emph{Computational Linguistics}}
  \bibinfo{volume}{41}, \bibinfo{number}{1} (\bibinfo{year}{2015}).
\newblock


\bibitem[Dubey et~al\mbox{.}(2024)]%
        {dubey2024llama}
\bibfield{author}{\bibinfo{person}{Abhimanyu Dubey}, \bibinfo{person}{Abhinav
  Jauhri}, \bibinfo{person}{Abhinav Pandey}, \bibinfo{person}{Abhishek Kadian},
  \bibinfo{person}{Ahmad Al-Dahle}, \bibinfo{person}{Aiesha Letman},
  \bibinfo{person}{Akhil Mathur}, \bibinfo{person}{Alan Schelten},
  \bibinfo{person}{Amy Yang}, \bibinfo{person}{Angela Fan}, {et~al\mbox{.}}}
  \bibinfo{year}{2024}\natexlab{}.
\newblock \showarticletitle{The llama 3 herd of models}.
\newblock \bibinfo{journal}{\emph{arXiv preprint arXiv:2407.21783}}
  (\bibinfo{year}{2024}).
\newblock


\bibitem[Fan et~al\mbox{.}(2019)]%
        {fan2019eli5}
\bibfield{author}{\bibinfo{person}{Angela Fan}, \bibinfo{person}{Yacine
  Jernite}, \bibinfo{person}{Ethan Perez}, \bibinfo{person}{David Grangier},
  \bibinfo{person}{Jason Weston}, {and} \bibinfo{person}{Michael Auli}.}
  \bibinfo{year}{2019}\natexlab{}.
\newblock \showarticletitle{ELI5: Long Form Question Answering}. In
  \bibinfo{booktitle}{\emph{Proceedings of the 57th Annual Meeting of the
  Association for Computational Linguistics}}. \bibinfo{pages}{3558--3567}.
\newblock


\bibitem[Fan et~al\mbox{.}(2025)]%
        {fan2025trustrag}
\bibfield{author}{\bibinfo{person}{Yixing Fan}, \bibinfo{person}{Qiang Yan},
  \bibinfo{person}{Wenshan Wang}, \bibinfo{person}{Jiafeng Guo},
  \bibinfo{person}{Ruqing Zhang}, {and} \bibinfo{person}{Xueqi Cheng}.}
  \bibinfo{year}{2025}\natexlab{}.
\newblock \showarticletitle{TrustRAG: An Information Assistant with Retrieval
  Augmented Generation}.
\newblock \bibinfo{journal}{\emph{arXiv preprint arXiv:2502.13719}}
  (\bibinfo{year}{2025}).
\newblock


\bibitem[Filippova(2020)]%
        {filippova2020controlled}
\bibfield{author}{\bibinfo{person}{Katja Filippova}.}
  \bibinfo{year}{2020}\natexlab{}.
\newblock \showarticletitle{Controlled Hallucinations: Learning to Generate
  Faithfully from Noisy Data}. In \bibinfo{booktitle}{\emph{Findings of the
  Association for Computational Linguistics: EMNLP 2020}}.
  \bibinfo{pages}{864--870}.
\newblock


\bibitem[Gao et~al\mbox{.}(2023a)]%
        {gao2023rarr}
\bibfield{author}{\bibinfo{person}{Luyu Gao}, \bibinfo{person}{Zhuyun Dai},
  \bibinfo{person}{Panupong Pasupat}, \bibinfo{person}{Anthony Chen},
  \bibinfo{person}{Arun~Tejasvi Chaganty}, \bibinfo{person}{Yicheng Fan},
  \bibinfo{person}{Vincent Zhao}, \bibinfo{person}{Ni Lao},
  \bibinfo{person}{Hongrae Lee}, \bibinfo{person}{Da-Cheng Juan},
  {et~al\mbox{.}}} \bibinfo{year}{2023}\natexlab{a}.
\newblock \showarticletitle{RARR: Researching and Revising What Language Models
  Say, Using Language Models}. In \bibinfo{booktitle}{\emph{Proceedings of the
  61st Annual Meeting of the Association for Computational Linguistics (Volume
  1: Long Papers)}}. \bibinfo{pages}{16477--16508}.
\newblock


\bibitem[Gao et~al\mbox{.}(2023c)]%
        {gao2023enabling}
\bibfield{author}{\bibinfo{person}{Tianyu Gao}, \bibinfo{person}{Howard Yen},
  \bibinfo{person}{Jiatong Yu}, {and} \bibinfo{person}{Danqi Chen}.}
  \bibinfo{year}{2023}\natexlab{c}.
\newblock \showarticletitle{Enabling Large Language Models to Generate Text
  with Citations}. In \bibinfo{booktitle}{\emph{Proceedings of the 2023
  Conference on Empirical Methods in Natural Language Processing}}.
  \bibinfo{pages}{6465--6488}.
\newblock


\bibitem[Gao et~al\mbox{.}(2023b)]%
        {gao2023retrieval}
\bibfield{author}{\bibinfo{person}{Yunfan Gao}, \bibinfo{person}{Yun Xiong},
  \bibinfo{person}{Xinyu Gao}, \bibinfo{person}{Kangxiang Jia},
  \bibinfo{person}{Jinliu Pan}, \bibinfo{person}{Yuxi Bi}, \bibinfo{person}{Yi
  Dai}, \bibinfo{person}{Jiawei Sun}, {and} \bibinfo{person}{Haofen Wang}.}
  \bibinfo{year}{2023}\natexlab{b}.
\newblock \showarticletitle{Retrieval-augmented generation for large language
  models: A survey}.
\newblock \bibinfo{journal}{\emph{arXiv preprint arXiv:2312.10997}}
  (\bibinfo{year}{2023}).
\newblock


\bibitem[GLM et~al\mbox{.}(2024)]%
        {glm2024chatglm}
\bibfield{author}{\bibinfo{person}{Team GLM}, \bibinfo{person}{Aohan Zeng},
  \bibinfo{person}{Bin Xu}, \bibinfo{person}{Bowen Wang},
  \bibinfo{person}{Chenhui Zhang}, \bibinfo{person}{Da Yin},
  \bibinfo{person}{Dan Zhang}, \bibinfo{person}{Diego Rojas},
  \bibinfo{person}{Guanyu Feng}, \bibinfo{person}{Hanlin Zhao},
  {et~al\mbox{.}}} \bibinfo{year}{2024}\natexlab{}.
\newblock \showarticletitle{Chatglm: A family of large language models from
  glm-130b to glm-4 all tools}.
\newblock \bibinfo{journal}{\emph{arXiv preprint arXiv:2406.12793}}
  (\bibinfo{year}{2024}).
\newblock


\bibitem[Guo et~al\mbox{.}(2019)]%
        {10.1145/3331184.3331403}
\bibfield{author}{\bibinfo{person}{Jiafeng Guo}, \bibinfo{person}{Yixing Fan},
  \bibinfo{person}{Xiang Ji}, {and} \bibinfo{person}{Xueqi Cheng}.}
  \bibinfo{year}{2019}\natexlab{}.
\newblock \showarticletitle{MatchZoo: A Learning, Practicing, and Developing
  System for Neural Text Matching}. In \bibinfo{booktitle}{\emph{Proceedings of
  the 42nd International ACM SIGIR Conference on Research and Development in
  Information Retrieval}} (Paris, France) \emph{(\bibinfo{series}{SIGIR'19})}.
  \bibinfo{publisher}{Association for Computing Machinery},
  \bibinfo{address}{New York, NY, USA}, \bibinfo{pages}{1297–1300}.
\newblock
\showISBNx{9781450361729}
\href{https://doi.org/10.1145/3331184.3331403}{doi:\nolinkurl{10.1145/3331184.3331403}}


\bibitem[Honovich et~al\mbox{.}(2022)]%
        {honovich2022true}
\bibfield{author}{\bibinfo{person}{Or Honovich}, \bibinfo{person}{Roee
  Aharoni}, \bibinfo{person}{Jonathan Herzig}, \bibinfo{person}{Hagai
  Taitelbaum}, \bibinfo{person}{Doron Kukliansy}, \bibinfo{person}{Vered
  Cohen}, \bibinfo{person}{Thomas Scialom}, \bibinfo{person}{Idan Szpektor},
  \bibinfo{person}{Avinatan Hassidim}, {and} \bibinfo{person}{Yossi Matias}.}
  \bibinfo{year}{2022}\natexlab{}.
\newblock \showarticletitle{TRUE: Re-evaluating Factual Consistency
  Evaluation}. In \bibinfo{booktitle}{\emph{Proceedings of the 2022 Conference
  of the North American Chapter of the Association for Computational
  Linguistics: Human Language Technologies}}. \bibinfo{pages}{3905--3920}.
\newblock


\bibitem[Hu et~al\mbox{.}(2024)]%
        {hu2024benchmarking}
\bibfield{author}{\bibinfo{person}{Nan Hu}, \bibinfo{person}{Jiaoyan Chen},
  \bibinfo{person}{Yike Wu}, \bibinfo{person}{Guilin Qi},
  \bibinfo{person}{Sheng Bi}, \bibinfo{person}{Tongtong Wu}, {and}
  \bibinfo{person}{Jeff~Z Pan}.} \bibinfo{year}{2024}\natexlab{}.
\newblock \showarticletitle{Benchmarking large language models in complex
  question answering attribution using knowledge graphs}.
\newblock \bibinfo{journal}{\emph{arXiv preprint arXiv:2401.14640}}
  (\bibinfo{year}{2024}).
\newblock


\bibitem[Huang and Chang(2023)]%
        {huang2023citation}
\bibfield{author}{\bibinfo{person}{Jie Huang} {and} \bibinfo{person}{Kevin
  Chen-Chuan Chang}.} \bibinfo{year}{2023}\natexlab{}.
\newblock \showarticletitle{Citation: A key to building responsible and
  accountable large language models}.
\newblock \bibinfo{journal}{\emph{arXiv preprint arXiv:2307.02185}}
  (\bibinfo{year}{2023}).
\newblock


\bibitem[Ji et~al\mbox{.}(2023)]%
        {ji2023survey}
\bibfield{author}{\bibinfo{person}{Ziwei Ji}, \bibinfo{person}{Nayeon Lee},
  \bibinfo{person}{Rita Frieske}, \bibinfo{person}{Tiezheng Yu},
  \bibinfo{person}{Dan Su}, \bibinfo{person}{Yan Xu}, \bibinfo{person}{Etsuko
  Ishii}, \bibinfo{person}{Ye~Jin Bang}, \bibinfo{person}{Andrea Madotto},
  {and} \bibinfo{person}{Pascale Fung}.} \bibinfo{year}{2023}\natexlab{}.
\newblock \showarticletitle{Survey of hallucination in natural language
  generation}.
\newblock \bibinfo{journal}{\emph{Comput. Surveys}} \bibinfo{volume}{55},
  \bibinfo{number}{12} (\bibinfo{year}{2023}), \bibinfo{pages}{1--38}.
\newblock


\bibitem[Lewis et~al\mbox{.}(2020)]%
        {lewis2020retrieval}
\bibfield{author}{\bibinfo{person}{Patrick Lewis}, \bibinfo{person}{Ethan
  Perez}, \bibinfo{person}{Aleksandra Piktus}, \bibinfo{person}{Fabio Petroni},
  \bibinfo{person}{Vladimir Karpukhin}, \bibinfo{person}{Naman Goyal},
  \bibinfo{person}{Heinrich K{\"u}ttler}, \bibinfo{person}{Mike Lewis},
  \bibinfo{person}{Wen-tau Yih}, \bibinfo{person}{Tim Rockt{\"a}schel},
  {et~al\mbox{.}}} \bibinfo{year}{2020}\natexlab{}.
\newblock \showarticletitle{Retrieval-augmented generation for
  knowledge-intensive nlp tasks}.
\newblock \bibinfo{journal}{\emph{Advances in Neural Information Processing
  Systems}}  \bibinfo{volume}{33} (\bibinfo{year}{2020}),
  \bibinfo{pages}{9459--9474}.
\newblock


\bibitem[Li et~al\mbox{.}(2024)]%
        {li2024improving}
\bibfield{author}{\bibinfo{person}{Dongfang Li}, \bibinfo{person}{Zetian Sun},
  \bibinfo{person}{Baotian Hu}, \bibinfo{person}{Zhenyu Liu},
  \bibinfo{person}{Xinshuo Hu}, \bibinfo{person}{Xuebo Liu}, {and}
  \bibinfo{person}{Min Zhang}.} \bibinfo{year}{2024}\natexlab{}.
\newblock \showarticletitle{Improving Attributed Text Generation of Large
  Language Models via Preference Learning}.
\newblock \bibinfo{journal}{\emph{arXiv preprint arXiv:2403.18381}}
  (\bibinfo{year}{2024}).
\newblock


\bibitem[Lin(2004)]%
        {lin2004rouge}
\bibfield{author}{\bibinfo{person}{Chin-Yew Lin}.}
  \bibinfo{year}{2004}\natexlab{}.
\newblock \showarticletitle{Rouge: A package for automatic evaluation of
  summaries}. In \bibinfo{booktitle}{\emph{Text summarization branches out}}.
  \bibinfo{pages}{74--81}.
\newblock


\bibitem[Liu et~al\mbox{.}(2024a)]%
        {liu2024deepseek}
\bibfield{author}{\bibinfo{person}{Aixin Liu}, \bibinfo{person}{Bei Feng},
  \bibinfo{person}{Bing Xue}, \bibinfo{person}{Bingxuan Wang},
  \bibinfo{person}{Bochao Wu}, \bibinfo{person}{Chengda Lu},
  \bibinfo{person}{Chenggang Zhao}, \bibinfo{person}{Chengqi Deng},
  \bibinfo{person}{Chenyu Zhang}, \bibinfo{person}{Chong Ruan},
  {et~al\mbox{.}}} \bibinfo{year}{2024}\natexlab{a}.
\newblock \showarticletitle{Deepseek-v3 technical report}.
\newblock \bibinfo{journal}{\emph{arXiv preprint arXiv:2412.19437}}
  (\bibinfo{year}{2024}).
\newblock


\bibitem[Liu et~al\mbox{.}(2023)]%
        {liu2023webglm}
\bibfield{author}{\bibinfo{person}{Xiao Liu}, \bibinfo{person}{Hanyu Lai},
  \bibinfo{person}{Hao Yu}, \bibinfo{person}{Yifan Xu}, \bibinfo{person}{Aohan
  Zeng}, \bibinfo{person}{Zhengxiao Du}, \bibinfo{person}{Peng Zhang},
  \bibinfo{person}{Yuxiao Dong}, {and} \bibinfo{person}{Jie Tang}.}
  \bibinfo{year}{2023}\natexlab{}.
\newblock \showarticletitle{WebGLM: Towards an efficient web-enhanced question
  answering system with human preferences}. In
  \bibinfo{booktitle}{\emph{Proceedings of the 29th ACM SIGKDD Conference on
  Knowledge Discovery and Data Mining}}. \bibinfo{pages}{4549--4560}.
\newblock


\bibitem[Liu et~al\mbox{.}(2024b)]%
        {liu2024invar}
\bibfield{author}{\bibinfo{person}{Ziwei Liu}, \bibinfo{person}{Liang Zhang},
  \bibinfo{person}{Qian Li}, \bibinfo{person}{Jianghua Wu}, {and}
  \bibinfo{person}{Guangxu Zhu}.} \bibinfo{year}{2024}\natexlab{b}.
\newblock \showarticletitle{Invar-RAG: Invariant LLM-aligned Retrieval for
  Better Generation}.
\newblock \bibinfo{journal}{\emph{arXiv preprint arXiv:2411.07021}}
  (\bibinfo{year}{2024}).
\newblock


\bibitem[Min et~al\mbox{.}(2020)]%
        {min2020ambigqa}
\bibfield{author}{\bibinfo{person}{Sewon Min}, \bibinfo{person}{Julian
  Michael}, \bibinfo{person}{Hannaneh Hajishirzi}, {and} \bibinfo{person}{Luke
  Zettlemoyer}.} \bibinfo{year}{2020}\natexlab{}.
\newblock \showarticletitle{AmbigQA: Answering ambiguous open-domain
  questions}.
\newblock \bibinfo{journal}{\emph{arXiv preprint arXiv:2004.10645}}
  (\bibinfo{year}{2020}).
\newblock


\bibitem[Nakano et~al\mbox{.}(2021)]%
        {nakano2021webgpt}
\bibfield{author}{\bibinfo{person}{Reiichiro Nakano}, \bibinfo{person}{Jacob
  Hilton}, \bibinfo{person}{Suchir Balaji}, \bibinfo{person}{Jeff Wu},
  \bibinfo{person}{Long Ouyang}, \bibinfo{person}{Christina Kim},
  \bibinfo{person}{Christopher Hesse}, \bibinfo{person}{Shantanu Jain},
  \bibinfo{person}{Vineet Kosaraju}, \bibinfo{person}{William Saunders},
  {et~al\mbox{.}}} \bibinfo{year}{2021}\natexlab{}.
\newblock \showarticletitle{Webgpt: Browser-assisted question-answering with
  human feedback}.
\newblock \bibinfo{journal}{\emph{arXiv preprint arXiv:2112.09332}}
  (\bibinfo{year}{2021}).
\newblock


\bibitem[Papineni et~al\mbox{.}(2002)]%
        {papineni2002bleu}
\bibfield{author}{\bibinfo{person}{Kishore Papineni}, \bibinfo{person}{Salim
  Roukos}, \bibinfo{person}{Todd Ward}, {and} \bibinfo{person}{Wei-Jing Zhu}.}
  \bibinfo{year}{2002}\natexlab{}.
\newblock \showarticletitle{Bleu: a method for automatic evaluation of machine
  translation}. In \bibinfo{booktitle}{\emph{Proceedings of the 40th annual
  meeting of the Association for Computational Linguistics}}.
  \bibinfo{pages}{311--318}.
\newblock


\bibitem[Press et~al\mbox{.}(2024)]%
        {press2024citeme}
\bibfield{author}{\bibinfo{person}{Ori Press}, \bibinfo{person}{Andreas
  Hochlehnert}, \bibinfo{person}{Ameya Prabhu}, \bibinfo{person}{Vishaal
  Udandarao}, \bibinfo{person}{Ofir Press}, {and} \bibinfo{person}{Matthias
  Bethge}.} \bibinfo{year}{2024}\natexlab{}.
\newblock \showarticletitle{Cite{ME}: Can Language Models Accurately Cite
  Scientific Claims?}. In \bibinfo{booktitle}{\emph{The Thirty-eight Conference
  on Neural Information Processing Systems Datasets and Benchmarks Track}}.
\newblock


\bibitem[Qin et~al\mbox{.}(2024)]%
        {qin2024large}
\bibfield{author}{\bibinfo{person}{Zhen Qin}, \bibinfo{person}{Rolf Jagerman},
  \bibinfo{person}{Kai Hui}, \bibinfo{person}{Honglei Zhuang},
  \bibinfo{person}{Junru Wu}, \bibinfo{person}{Le Yan},
  \bibinfo{person}{Jiaming Shen}, \bibinfo{person}{Tianqi Liu},
  \bibinfo{person}{Jialu Liu}, \bibinfo{person}{Donald Metzler},
  {et~al\mbox{.}}} \bibinfo{year}{2024}\natexlab{}.
\newblock \showarticletitle{Large Language Models are Effective Text Rankers
  with Pairwise Ranking Prompting}. In \bibinfo{booktitle}{\emph{Findings of
  the Association for Computational Linguistics: NAACL 2024}}.
  \bibinfo{pages}{1504--1518}.
\newblock


\bibitem[Rashkin et~al\mbox{.}(2023)]%
        {rashkin2023measuring}
\bibfield{author}{\bibinfo{person}{Hannah Rashkin}, \bibinfo{person}{Vitaly
  Nikolaev}, \bibinfo{person}{Matthew Lamm}, \bibinfo{person}{Lora Aroyo},
  \bibinfo{person}{Michael Collins}, \bibinfo{person}{Dipanjan Das},
  \bibinfo{person}{Slav Petrov}, \bibinfo{person}{Gaurav~Singh Tomar},
  \bibinfo{person}{Iulia Turc}, {and} \bibinfo{person}{David Reitter}.}
  \bibinfo{year}{2023}\natexlab{}.
\newblock \showarticletitle{Measuring Attribution in Natural Language
  Generation Models}.
\newblock \bibinfo{journal}{\emph{Computational Linguistics}}
  \bibinfo{volume}{49}, \bibinfo{number}{4} (\bibinfo{year}{2023}),
  \bibinfo{pages}{777--840}.
\newblock


\bibitem[Sachan et~al\mbox{.}(2022)]%
        {sachan2022improving}
\bibfield{author}{\bibinfo{person}{Devendra Sachan}, \bibinfo{person}{Mike
  Lewis}, \bibinfo{person}{Mandar Joshi}, \bibinfo{person}{Armen Aghajanyan},
  \bibinfo{person}{Wen-tau Yih}, \bibinfo{person}{Joelle Pineau}, {and}
  \bibinfo{person}{Luke Zettlemoyer}.} \bibinfo{year}{2022}\natexlab{}.
\newblock \showarticletitle{Improving Passage Retrieval with Zero-Shot Question
  Generation}. In \bibinfo{booktitle}{\emph{Proceedings of the 2022 Conference
  on Empirical Methods in Natural Language Processing}}.
  \bibinfo{pages}{3781--3797}.
\newblock


\bibitem[Schulman et~al\mbox{.}(2017)]%
        {schulman2017proximal}
\bibfield{author}{\bibinfo{person}{John Schulman}, \bibinfo{person}{Filip
  Wolski}, \bibinfo{person}{Prafulla Dhariwal}, \bibinfo{person}{Alec Radford},
  {and} \bibinfo{person}{Oleg Klimov}.} \bibinfo{year}{2017}\natexlab{}.
\newblock \showarticletitle{Proximal policy optimization algorithms}.
\newblock \bibinfo{journal}{\emph{arXiv preprint arXiv:1707.06347}}
  (\bibinfo{year}{2017}).
\newblock


\bibitem[Shuster et~al\mbox{.}(2021)]%
        {shuster2021retrieval}
\bibfield{author}{\bibinfo{person}{Kurt Shuster}, \bibinfo{person}{Spencer
  Poff}, \bibinfo{person}{Moya Chen}, \bibinfo{person}{Douwe Kiela}, {and}
  \bibinfo{person}{Jason Weston}.} \bibinfo{year}{2021}\natexlab{}.
\newblock \showarticletitle{Retrieval Augmentation Reduces Hallucination in
  Conversation}. In \bibinfo{booktitle}{\emph{Findings of the Association for
  Computational Linguistics: EMNLP 2021}}. \bibinfo{pages}{3784--3803}.
\newblock


\bibitem[Stelmakh et~al\mbox{.}(2022)]%
        {stelmakh2022asqa}
\bibfield{author}{\bibinfo{person}{Ivan Stelmakh}, \bibinfo{person}{Yi Luan},
  \bibinfo{person}{Bhuwan Dhingra}, {and} \bibinfo{person}{Ming-Wei Chang}.}
  \bibinfo{year}{2022}\natexlab{}.
\newblock \showarticletitle{ASQA: Factoid Questions Meet Long-Form Answers}. In
  \bibinfo{booktitle}{\emph{Proceedings of the 2022 Conference on Empirical
  Methods in Natural Language Processing}}. \bibinfo{pages}{8273--8288}.
\newblock


\bibitem[Sun et~al\mbox{.}(2024)]%
        {sun2024towards}
\bibfield{author}{\bibinfo{person}{Hao Sun}, \bibinfo{person}{Hengyi Cai},
  \bibinfo{person}{Bo Wang}, \bibinfo{person}{Yingyan Hou},
  \bibinfo{person}{Xiaochi Wei}, \bibinfo{person}{Shuaiqiang Wang},
  \bibinfo{person}{Yan Zhang}, {and} \bibinfo{person}{Dawei Yin}.}
  \bibinfo{year}{2024}\natexlab{}.
\newblock \showarticletitle{Towards Verifiable Text Generation with Evolving
  Memory and Self-Reflection}. In \bibinfo{booktitle}{\emph{Proceedings of the
  2024 Conference on Empirical Methods in Natural Language Processing}}.
  \bibinfo{pages}{8211--8227}.
\newblock


\bibitem[Sun et~al\mbox{.}(2023)]%
        {sun2023chatgpt}
\bibfield{author}{\bibinfo{person}{Weiwei Sun}, \bibinfo{person}{Lingyong Yan},
  \bibinfo{person}{Xinyu Ma}, \bibinfo{person}{Shuaiqiang Wang},
  \bibinfo{person}{Pengjie Ren}, \bibinfo{person}{Zhumin Chen},
  \bibinfo{person}{Dawei Yin}, {and} \bibinfo{person}{Zhaochun Ren}.}
  \bibinfo{year}{2023}\natexlab{}.
\newblock \showarticletitle{Is ChatGPT Good at Search? Investigating Large
  Language Models as Re-Ranking Agents}. In
  \bibinfo{booktitle}{\emph{Proceedings of the 2023 Conference on Empirical
  Methods in Natural Language Processing}}. \bibinfo{pages}{14918--14937}.
\newblock


\bibitem[Team et~al\mbox{.}(2024)]%
        {team2024gemma}
\bibfield{author}{\bibinfo{person}{Gemma Team}, \bibinfo{person}{Morgane
  Riviere}, \bibinfo{person}{Shreya Pathak}, \bibinfo{person}{Pier~Giuseppe
  Sessa}, \bibinfo{person}{Cassidy Hardin}, \bibinfo{person}{Surya
  Bhupatiraju}, \bibinfo{person}{L{\'e}onard Hussenot}, \bibinfo{person}{Thomas
  Mesnard}, \bibinfo{person}{Bobak Shahriari}, \bibinfo{person}{Alexandre
  Ram{\'e}}, {et~al\mbox{.}}} \bibinfo{year}{2024}\natexlab{}.
\newblock \showarticletitle{Gemma 2: Improving open language models at a
  practical size}.
\newblock \bibinfo{journal}{\emph{arXiv preprint arXiv:2408.00118}}
  (\bibinfo{year}{2024}).
\newblock


\bibitem[Thakur et~al\mbox{.}(2023)]%
        {thakur2023nomiracl}
\bibfield{author}{\bibinfo{person}{Nandan Thakur}, \bibinfo{person}{Luiz
  Bonifacio}, \bibinfo{person}{Xinyu Zhang}, \bibinfo{person}{Odunayo
  Ogundepo}, \bibinfo{person}{Ehsan Kamalloo}, \bibinfo{person}{David
  Alfonso-Hermelo}, \bibinfo{person}{Xiaoguang Li}, \bibinfo{person}{Qun Liu},
  \bibinfo{person}{Boxing Chen}, \bibinfo{person}{Mehdi Rezagholizadeh},
  {et~al\mbox{.}}} \bibinfo{year}{2023}\natexlab{}.
\newblock \showarticletitle{NoMIRACL: Knowing When You Don't Know for Robust
  Multilingual Retrieval-Augmented Generation}.
\newblock \bibinfo{journal}{\emph{arXiv preprint arXiv:2312.11361}}
  (\bibinfo{year}{2023}).
\newblock


\bibitem[Trivedi et~al\mbox{.}(2022)]%
        {trivedi2022musique}
\bibfield{author}{\bibinfo{person}{Harsh Trivedi}, \bibinfo{person}{Niranjan
  Balasubramanian}, \bibinfo{person}{Tushar Khot}, {and}
  \bibinfo{person}{Ashish Sabharwal}.} \bibinfo{year}{2022}\natexlab{}.
\newblock \showarticletitle{MuSiQue: Multi-hop Questions via Single-hop
  Question Composition}.
\newblock \bibinfo{journal}{\emph{Transactions of the Association for
  Computational Linguistics}}  \bibinfo{volume}{10} (\bibinfo{year}{2022}),
  \bibinfo{pages}{539--554}.
\newblock


\bibitem[Trivedi et~al\mbox{.}(2023)]%
        {trivedi2023interleaving}
\bibfield{author}{\bibinfo{person}{Harsh Trivedi}, \bibinfo{person}{Niranjan
  Balasubramanian}, \bibinfo{person}{Tushar Khot}, {and}
  \bibinfo{person}{Ashish Sabharwal}.} \bibinfo{year}{2023}\natexlab{}.
\newblock \showarticletitle{Interleaving Retrieval with Chain-of-Thought
  Reasoning for Knowledge-Intensive Multi-Step Questions}. In
  \bibinfo{booktitle}{\emph{Proceedings of the 61st Annual Meeting of the
  Association for Computational Linguistics (Volume 1: Long Papers)}}.
  \bibinfo{pages}{10014--10037}.
\newblock


\bibitem[Xia et~al\mbox{.}(2015)]%
        {10.1145/2766462.2767710}
\bibfield{author}{\bibinfo{person}{Long Xia}, \bibinfo{person}{Jun Xu},
  \bibinfo{person}{Yanyan Lan}, \bibinfo{person}{Jiafeng Guo}, {and}
  \bibinfo{person}{Xueqi Cheng}.} \bibinfo{year}{2015}\natexlab{}.
\newblock \showarticletitle{Learning Maximal Marginal Relevance Model via
  Directly Optimizing Diversity Evaluation Measures}. In
  \bibinfo{booktitle}{\emph{Proceedings of the 38th International ACM SIGIR
  Conference on Research and Development in Information Retrieval}} (Santiago,
  Chile) \emph{(\bibinfo{series}{SIGIR '15})}. \bibinfo{publisher}{Association
  for Computing Machinery}, \bibinfo{address}{New York, NY, USA},
  \bibinfo{pages}{113–122}.
\newblock
\showISBNx{9781450336215}
\href{https://doi.org/10.1145/2766462.2767710}{doi:\nolinkurl{10.1145/2766462.2767710}}


\bibitem[Yang et~al\mbox{.}(2024)]%
        {yang2024qwen2}
\bibfield{author}{\bibinfo{person}{An Yang}, \bibinfo{person}{Baosong Yang},
  \bibinfo{person}{Beichen Zhang}, \bibinfo{person}{Binyuan Hui},
  \bibinfo{person}{Bo Zheng}, \bibinfo{person}{Bowen Yu},
  \bibinfo{person}{Chengyuan Li}, \bibinfo{person}{Dayiheng Liu},
  \bibinfo{person}{Fei Huang}, \bibinfo{person}{Haoran Wei}, {et~al\mbox{.}}}
  \bibinfo{year}{2024}\natexlab{}.
\newblock \showarticletitle{Qwen2. 5 technical report}.
\newblock \bibinfo{journal}{\emph{arXiv preprint arXiv:2412.15115}}
  (\bibinfo{year}{2024}).
\newblock


\bibitem[Yang et~al\mbox{.}(2018)]%
        {yang2018hotpotqa}
\bibfield{author}{\bibinfo{person}{Zhilin Yang}, \bibinfo{person}{Peng Qi},
  \bibinfo{person}{Saizheng Zhang}, \bibinfo{person}{Yoshua Bengio},
  \bibinfo{person}{William Cohen}, \bibinfo{person}{Ruslan Salakhutdinov},
  {and} \bibinfo{person}{Christopher~D Manning}.}
  \bibinfo{year}{2018}\natexlab{}.
\newblock \showarticletitle{HotpotQA: A Dataset for Diverse, Explainable
  Multi-hop Question Answering}. In \bibinfo{booktitle}{\emph{Proceedings of
  the 2018 Conference on Empirical Methods in Natural Language Processing}}.
  \bibinfo{pages}{2369--2380}.
\newblock


\bibitem[Yin et~al\mbox{.}(2021)]%
        {yin2021docnli}
\bibfield{author}{\bibinfo{person}{Wenpeng Yin}, \bibinfo{person}{Dragomir
  Radev}, {and} \bibinfo{person}{Caiming Xiong}.}
  \bibinfo{year}{2021}\natexlab{}.
\newblock \showarticletitle{DocNLI: A Large-scale Dataset for Document-level
  Natural Language Inference}. In \bibinfo{booktitle}{\emph{Findings of the
  Association for Computational Linguistics: ACL-IJCNLP 2021}}.
  \bibinfo{pages}{4913--4922}.
\newblock


\bibitem[Zhang et~al\mbox{.}(2024)]%
        {zhang2024longcite}
\bibfield{author}{\bibinfo{person}{Jiajie Zhang}, \bibinfo{person}{Yushi Bai},
  \bibinfo{person}{Xin Lv}, \bibinfo{person}{Wanjun Gu},
  \bibinfo{person}{Danqing Liu}, \bibinfo{person}{Minhao Zou},
  \bibinfo{person}{Shulin Cao}, \bibinfo{person}{Lei Hou},
  \bibinfo{person}{Yuxiao Dong}, \bibinfo{person}{Ling Feng}, {and}
  \bibinfo{person}{Juanzi Li}.} \bibinfo{year}{2024}\natexlab{}.
\newblock \showarticletitle{LongCite: Enabling LLMs to Generate Fine-grained
  Citations in Long-context QA}.
\newblock \bibinfo{journal}{\emph{arXiv preprint arXiv:2409.02897}}
  (\bibinfo{year}{2024}).
\newblock


\end{thebibliography}

\appendix

\section{Prompt}
\subsection{Initial Answer Generation}
\label{A:1}

\begin{tcolorbox}
[title=Prompt Template for Initial Answer Generation,fontupper=\ttfamily,colback=gray!10!white,colframe=black,arc=1mm,boxrule=1pt,left=1mm,right=1mm,top=1mm,bottom=1mm, fonttitle=\small]
\small
Instruction: Please refer to the information in the following passages to answer the question. When answering, ignore any irrelevant information from the passages, but retain all relevant details to provide a comprehensive and accurate response. Always cite for any factual claim. When citing several search results, use [1][2][3]. Cite at least one passage in each sentence.

\textbf{Question}: \{Question\}

\textbf{Document}: [1](Title: \{Title\}): \{Passage\}

\textbf{Document}: [2](Title: \{Title\}): \{Passage\}

...

\textbf{Answer}:
\end{tcolorbox}

\subsection{Supporting Evidence Check}
\label{A:2}

\begin{tcolorbox}
[title=Prompt Template for Supporting Evidence Check,fontupper=\ttfamily,colback=gray!10!white,colframe=black,arc=1mm,boxrule=1pt,left=1mm,right=1mm,top=1mm,bottom=1mm, fonttitle=\small]
\small
Instruction: Please refer to the information in the following passage to answer the question. You need to first determine whether the information in the passage is helpful for answering the question. If you believe the passage is helpful, output 'Yes'; otherwise, output 'No'. Do not output any additional content.

\textbf{Question}: \{Question\}

\textbf{Passage}: \{Passage\}

\textbf{Response}:
\end{tcolorbox}

\subsection{Supporting Evidence Extraction}
\label{A:3}

\begin{tcolorbox}
[title=Prompt Template for Supporting Evidence Extraction,fontupper=\ttfamily,colback=gray!10!white,colframe=black,arc=1mm,boxrule=1pt,left=1mm,right=1mm,top=1mm,bottom=1mm, fonttitle=\small]
\small
Instruction: Please refer to the information in the following passage to answer the question. When answering, ignore any irrelevant information from the passage, but retain all relevant details to provide a comprehensive and accurate response.

\textbf{Question}: \{Question\}

\textbf{Passage}: \{Passage\}

\textbf{Response}:
\end{tcolorbox}

\subsection{Final Answer Refinement}
\label{A:4}

\begin{tcolorbox}
[title=Prompt Template for Final Answer Refinement,fontupper=\ttfamily,colback=gray!10!white,colframe=black,arc=1mm,boxrule=1pt,left=1mm,right=1mm,top=1mm,bottom=1mm, fonttitle=\small]
\small
Instruction: Please answer the following question. I will provide you with some answer statements with citations, as well as their original references. You need to summarize these statements and merge their citations such as [1][2].

\textbf{Question}: \{Question\}

\textbf{References}:

\textbf{Document}: [1](Title: \{Title\}): \{Passage\}

\textbf{Document}: [2](Title: \{Title\}): \{Passage\}

...

\textbf{Answer statements}:

\{Statement 1\} [citation ids]

\{Statement 2\} [citation ids]

...

\textbf{Your Answer}:

\end{tcolorbox}

\end{document}